\documentclass[12pt,a4paper]{article}
\usepackage[english]{babel}
\usepackage{graphicx}

\newcommand{\alt}{\mathbin{\lower 3pt\hbox
   {$\rlap{\raise 5pt\hbox{$\char'074$}}\mathchar"7218$}}}
\newcommand{\agt}{\mathbin{\lower 3pt\hbox
   {$\rlap{\raise 5pt\hbox{$\char'076$}}\mathchar"7218$}}}

\begin{document}
\setcounter{footnote}{0}
\setcounter{equation}{0}
\setcounter{figure}{0}
\setcounter{table}{0}
\vspace*{5mm}

\begin{center}
{\large\bf Is  $\varphi^4$ theory trivial ?
}

\vspace{4mm}
\vspace{4mm}
I. M. Suslov \\
P.L.Kapitza Institute for Physical Problems,
\\ 119337 Moscow, Russia \\
\vspace{1mm}
\end{center}

\begin{center}
\begin{minipage}{135mm}
{\bf Abstract } \\
The four-dimensional $\varphi^4$ theory is usually considered
to be trivial in the continuum limit. In fact,  two
definitions of triviality were mixed in the literature.
The first one, introduced by Wilson, is equivalent to
positiveness of the Gell-Mann -- Low function $\beta(g)$ for
$g\ne 0$; it is confirmed by all available information and can be
considered as firmly established. The second definition,
introduced by mathematical community, corresponds to the true
triviality, i.e.  principal impossibility to construct continuous
theory with finite interaction at large distances: it needs not
only positiveness of $\beta(g)$ but also its
sufficiently quick growth at infinity.  Indications of true
triviality are not numerous and allow different interpretation.
According to the recent results, such triviality is surely
absent.
\end{minipage}
\end{center}


\vspace{6mm}
\begin{center}
{\bf 1. Introduction}
\end{center}

The problem of the "zero charge" or "triviality" of quantum
field theories was raised firstly by Landau and co-workers
\cite{1,3}. According to Landau, Abrikosov, Khalatnikov
\cite{1}, relation of the bare charge $g_0$ with observable
charge $g$ for renormalizable field theories is
given by expression
$$
g=\frac{g_0}{1+\beta_2 g_0 \ln \Lambda^2/m^2} \,,
\eqno(1)
$$
where $m$ is the mass of the particle, and $\Lambda$ is
the momentum cut-off.  For finite $g_0$ and
$\Lambda\to \infty$,  the observable charge  $g\to 0$ and
the "zero charge" situation takes place. The proper
interpretation of Eq.1 consists in its inverting,
$$
g_0=\frac{g}{1-\beta_2 g \ln \Lambda^2/m^2} \,,
\eqno(2)
$$
so that the bare charge $g_0$ is related to the length
scale $\Lambda^{-1}$ and chosen to give a correct value of $g$.
The growth of $g_0$ with  $\Lambda$ invalidates Eqs.1,2
in the region $g_0\sim 1$ and existence of the "Landau pole" in
Eq.2 has no physical sense.

The actual behavior of the charge  $g(L)$ as a function of the
length  scale $L$ is determined by the Gell-Mann -- Low equation
$$
-\frac{dg}{d \ln L^2} =\beta(g)=\beta_2 g^2+\beta_3 g^3+\ldots
\eqno(3)
$$
and depends on appearance of the function  $\beta(g)$. According
to classification by Bogolyubov and Shirkov \cite{2}, the growth
of  $g(L)$ is saturated, if $\beta(g)$ has a zero for finite  $g$,
and continues to infinity, if $\beta(g)$ is non-alternating and
behaves as $\beta(g)\sim g^\alpha$ with $\alpha\le 1$ for large
$g$; if, however, $\beta(g)\sim g^\alpha$ with $\alpha>1$, then
$g(L) $ is divergent  at finite $L=L_0$  (the real Landau pole
arises) and the theory is internally inconsistent due to
indeterminacy of $g(L)$ for $L<L_0$.
Landau and Pomeranchuk \cite{3} tried to justify the latter
possibility, arguing that Eq.1 is valid without restrictions;
however, it is possible only for the strict equality
$\beta(g)=\beta_2 g^2$, which is surely invalid due to finiteness
of $\beta_3$.

One can see that solution of the "zero charge" problem
needs calculation of the Gell-Mann -- Low function
$\beta(g)$ at arbitrary $g$, and in particular its asymptotic
behavior for $g\to\infty$. This problem is very difficult and
corresponding information has appeared only recently (Sec.4).
Nevertheless, scientific community looks rather convinced
in triviality of $\varphi^4$ theory \cite{21}--\cite{54}.
Such situation is rather
strange, since attempts to study strong coupling behavior
of quantum field theories are not numerous and their results
cannot be considered as commonly accepted.

In fact, two definitions of triviality were mixed in the literature.
The first one, introduced by Wilson
\cite{21}
(Sec.2), is
equivalent to positiveness of $\beta(g)$ for $g\ne 0$; it is
confirmed by all available information and can be considered as
firmly established. The second definition, introduced by
mathematical community
\cite{29,30,31}  (Sec.3), corresponds to the true
triviality and is equivalent to internal inconsistency in the
Bogolyubov and Shirkov sense: it
requires not only positiveness of
$\beta(g)$ but also corresponding asymptotical behavior. Evidence
of true triviality is not extensive and allows different
interpretation (Sec.5,6): according to recent results (Sec.4)
such triviality is absent. These recent results \cite{0}  give new
insight to the problem: to obtain nontrivial theory one needs to
use the complex values of the bare charge $g_0$, which were never
exploited in mathematical proofs and numerical simulations.

In what follows, we have in mind the $O(n)$--symmetric
$\varphi^4$ theory with an action
$$
S\{\varphi\} =\int \,d^dx \left\{
{\textstyle\frac{1}{2}}(\nabla\varphi)^2
+ {\textstyle\frac{1}{2}} m^2 \varphi^{\,2} +
{\textstyle\frac{1}{8}} u
\varphi^{\,4} \right\}\,,
\eqno(4)
$$
$$
u=g_0\Lambda^{\epsilon}\,, \qquad \epsilon=4-d
$$
in $d$--dimensional space.

\vspace{3mm}
\begin{center}
{\bf 2. Triviality in Wilson's sense}
\end{center}
\vspace{2mm}

In the theory of critical phenomena, Eq.1 has entirely different
interpretation. In this case, the cut-off $\Lambda$ and the bare
charge  $g_0$ have a direct physical sense and are related with
a lattice spacing and the coefficient in the effective Landau
Hamiltonian. The "zero charge" situation occurs in this case for
$m\to 0$, i.e. at approaching the phase transition point, and
corresponds to the absence of interaction between large-scale
fluctuations of the order parameter. According to Wilson's
renormalization group analysis \cite{21a}, the $\varphi^4$ theory
reduces at large distances  to the trivial Gaussian model for
space dimensionality $d\ge 4$. Success of Wilson's
$\epsilon$--expansion \cite{21a,22,16} is directly related with
this triviality: for $d=4-\epsilon$, interaction between
large-scale fluctuations becomes finite but small for $\epsilon\ll
1$.

In subsequent papers, Wilson set problem more deeply: does
triviality for $d=4$ exist only for small $g_0$, or has the global
character? The answer depends on the properties of the
$\beta$-function: if $\beta(g)$ has no non-trivial roots
(Fig.\,1,a),
\begin{figure}
\centerline{\includegraphics[width=5.1 in]{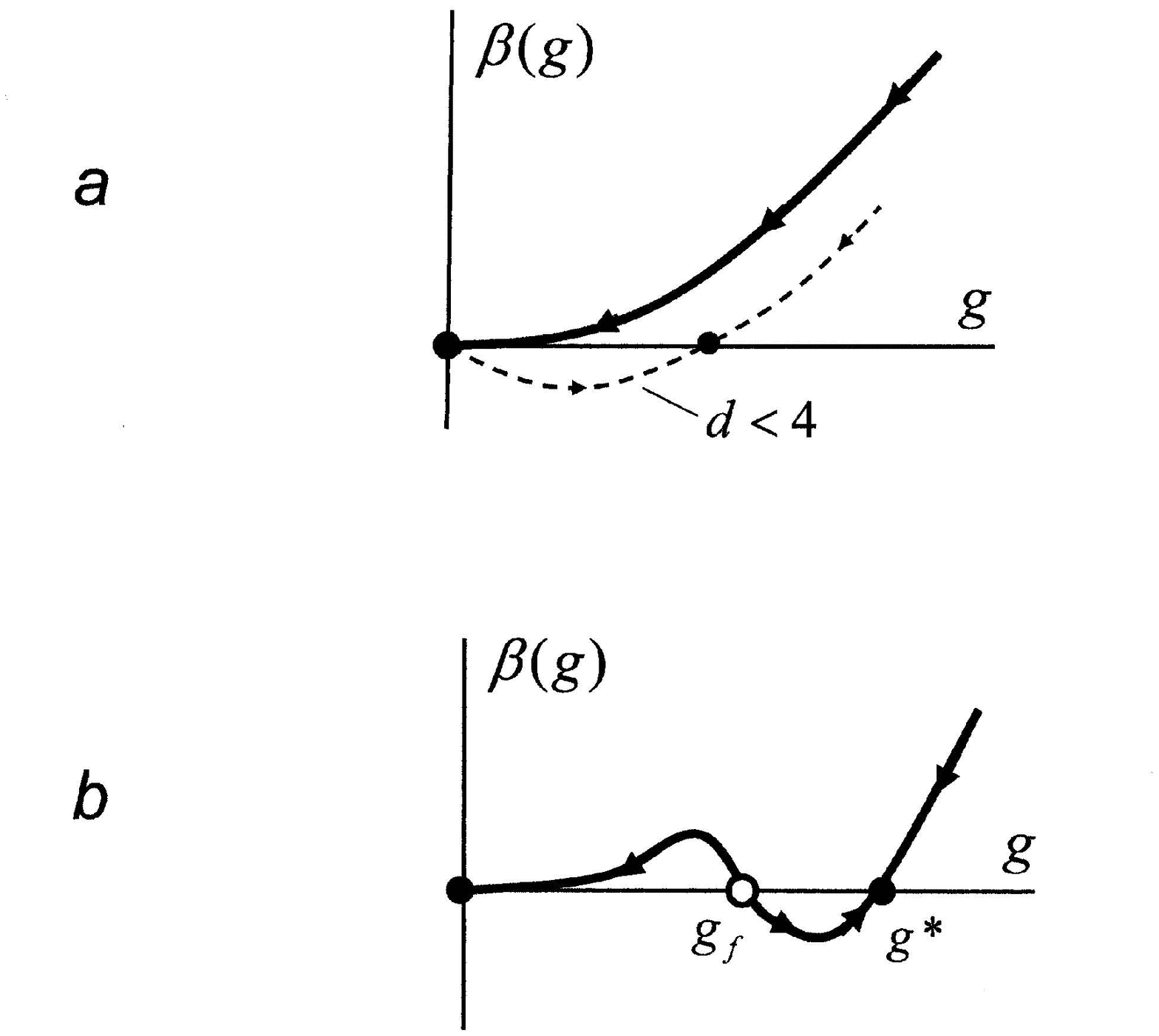}}
\caption{Flow of $g$ with increase in $L$ according to the
Gell-Mann -- Low equation : (a) in the case of non-alternating
$\beta(g)$, evolution ends in the Gaussian fixed point $g=0$; (b)
in the case of alternating $\beta(g)$, the domain of attraction of
the Gaussian fixed point is restricted by the boundary $g_f$. For
$d<4$, $\beta$-function has a negative portion (dashed line in
Fig.1,a).} \label{fig1}
\end{figure}
then effective interaction tends to zero
at large distances for any initial value $g_0$. If, however,
$\beta(g)$ is alternating  (Fig.\,1,b), then non-trivial limit
$g^*$ may occur at large length scales. The latter possibility is
of essential interest for the condensed matter physics
\cite{201}:  it means existence of phase transitions of the new
type, which are not described by Wilson's  $\epsilon$--expansion.

Using logic of proof by contradiction, Wilson assumed
existence of the boundary  $g_f$ for the domain of attraction
of the Gaussian fixed point  $g=0$ (which is equivalent to
alternating behavior for  $\beta(g)$) and derived the consequences
convenient for numerical verification. According to his results
\cite{21}, there are no indications of existence $g_f$.
Historically, it was the first real attempt to
investigate the strong coupling region for  $\varphi^4$ theory
and the first evidence of non-alternating character of $\beta(g)$.

\vspace{3mm}
\begin{center}
{\bf 3. Triviality in mathematical sense.}
\end{center}
\vspace{2mm}

Another definition of triviality was given in the mathematical
papers  \cite{29}--\cite{31}. If a field theory is understood as
a limit of lattice theories, then one can introduce the bare
charge $g_0$ as a function of interatomic spacing  $a_0$.
A theory is nontrivial, if for some choice of dependence
 $g_0(a_0)$ one can take the limit $a_0\to 0$ and provide
 finite interaction at large distances; if it is impossible
 for any choice of $g_0(a_0)$, then a theory is trivial.
Such definition corresponds to the true triviality,
i.e. principal impossibility to construct continuous theory
with finite interaction at large $L$. It is equivalent to
internal inconsistency in the Bogolyubov and Shirkov sense
(Sec.1). Indeed, in the latter case  a theory does not exist
for scales $L<L_0$, if a charge $g_\infty$ is finite for
$L\agt m^{-1}$; realization of the limit  $a_0\to 0$ demands
to diminish $L_0$ till zero, which is possible only for
 $g_\infty\to 0$.

It was rigorously proved in \cite{29}--\cite{31} that $\varphi^4$
theory is trivial for  $d>4$ and nontrivial for $d<4$; using
experience of these proofs, some plausible arguments were given in
favor of triviality for $d=4$. From the physical viewpoint, these
results are rather evident. Indeed, $\varphi^4$ theory is
nonrenormalizable for $d>4$ and the limit  $a_0\to 0$ cannot be
taken without destroying its structure;
in the given definition of triviality, the structure of
$\varphi^4$ theory is maintained artificially for arbitrary small
$a_0$, and hence the only possibility for it is to "throw off"
interaction and transfer to the Gaussian theory.  Non-triviality
of $\varphi^4$ theory for $d<4$ is related with the negative
portion of the $\beta$-function  (Fig.\,1,a, dashed line), for
which $g(L)\to g^*$ at large distances and $g(L)\to 0$ for  $L\to
0$; existence of this negative portion  can be demonstrated
analytically for $d=4-\epsilon$ with $\epsilon\ll 1$ and
numerically for $d=2$ and  $d=3$ \cite{16a}. One can see, that the
results proved in \cite{29}--\cite{31} do not require any study of
the strong coupling region, and hence no propositions can be made
for the case $d=4$, where such investigation is obligatory. In
fact, to obtain nontrivial theory for $d=4$, one needs to use the
complex values of $g_0$ (Sec.4), which were never considered in
mathematical proofs.  \vspace{2mm}

\qquad\qquad\qquad\qquad\qquad\qquad
--------------------------------
\vspace{2mm}

Above discussion makes clear the difference between two
definitions of triviality. Triviality in Wilson's sense needs
only positiveness of the $\beta$--function for $g\ne 0$, while the
true triviality demands in addition its sufficiently quick growth
at large $g$,  $\beta(g)\sim g^\alpha$ with $\alpha>1$.
This difference is practically not understood in the
literature. Some authors (see e.g. \cite{34,46}) clearly
state that the limits $\Lambda\to\infty$ and $m\to 0$
are equivalent. Indeed, the formal solution of Eq.3
$$
\int\limits_{g_m}^{g_\Lambda} \,\frac{dg}{\beta(g)}=
\ln\frac{\Lambda^2}{m^2}
\eqno(5)
$$
is determined only by the ratio  $\Lambda/m$;
however, its physical consequences depend on setting
the problem. If  $\Lambda$ and $g_\Lambda$ are fixed, then
for positive $\beta(g)$ we always have  $g_m\to 0$ for $m\to 0$.
If $m$ and  $g_m$ are fixed, then the possibility
$g_\Lambda\to\infty$, $\Lambda\to\infty$ is realized
only for $\alpha\le 1$, while in the opposite case the limit
$\Lambda\to\infty$ is impossible at all.

\vspace{3mm}
\begin{center}
{\bf 4. Available information on the $\beta$-function for
$d=4$.  }
\end{center}
\vspace{2mm}

Information on the $\beta$-function in $\varphi^4$ theory
can be obtained using the fact that
the first four coefficients $\beta_N$
in Eq.3 are known from diagrammatic calculations
\cite{7,8}, while their large order behavior
$$
\beta_N^{as} =c a^N \Gamma(N+b)
\eqno(6)
$$
can be established by the Lipatov method \cite{10,9}. Smooth
interpolation of the coefficient function and the proper summation
of the perturbation series allows in principle to obtain
$\beta(g)$ for all $g$. The general appearance of the
$\beta$-function in the four-dimensional $\varphi^4$ theory,
obtained in \cite{4}, is shown in Fig.2, as well as the results of
some other authors \cite{26}-\cite{28}.
\begin{figure}
\centerline{\includegraphics[width=6.5 in]{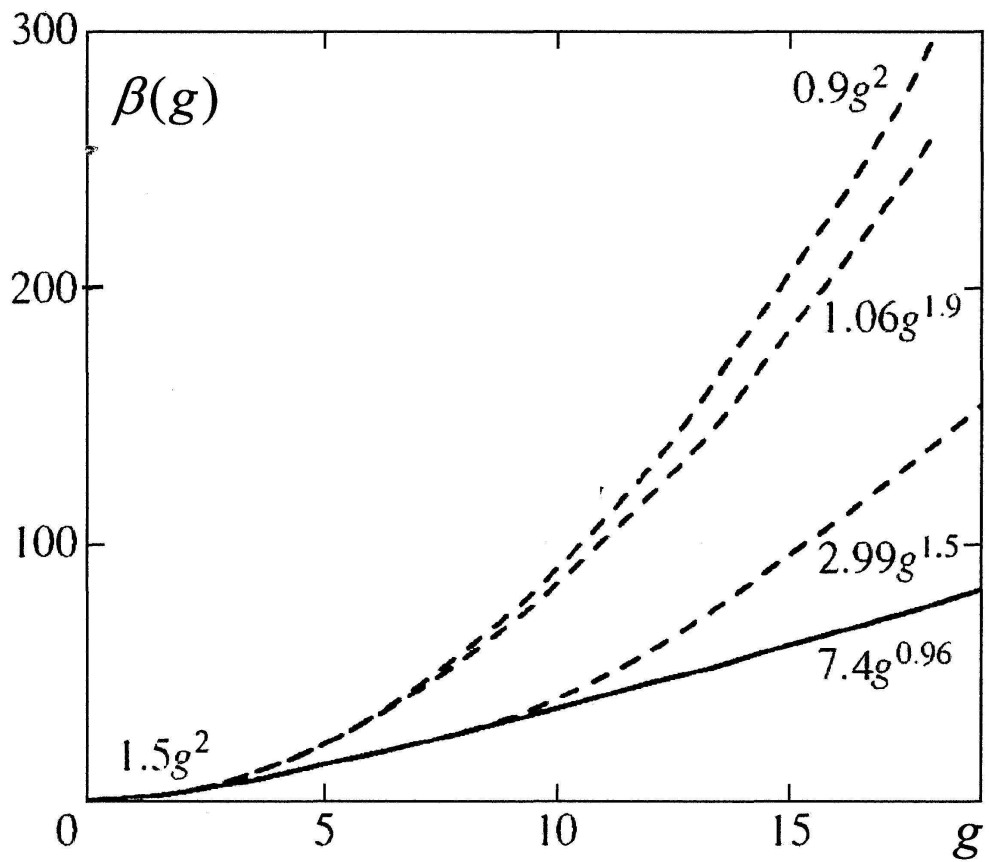}}
\caption{General appearance of the Gell-Mann -- Low function in
four-dimensional $\varphi^4$ theory according to \cite{4} (solid
line) and the results of other authors (dashed lines from top to
bottom correspond to papers \cite{26}, \cite{27}, \cite{28}). }
\label{fig2}
\end{figure}
There is no doubt that $\beta(g)$ is positive and hence
triviality in Wilson's sense does exist. There are also
grounds to expect manifestations of true triviality.
Indeed, Fig.2 corresponds to the "natural" normalization of
charge, when parameter  $a$ in the Lipatov asymptotics (6) is
equal to unity, while the interaction term in the action (4) is
written as  $(16\pi^2/4!)g\varphi^4$.   In this case, the nearest
singularity in the Borel plane lies at the unit distance from the
origin, and $\beta(g)$ is expected to change on the scale of the
order of unity. It is more or less so (Fig.2), but the one-loop
behavior appears to be somewhat dragged-out:  approximately
quadratic dependence continues till $g\sim 10$. For the
conventional normalization of charge, when the interaction term is
written as $g\varphi^4/8$ or $g\varphi^4/4!$,  the boundary
between "weak coupling" and "strong coupling" regions lies at
$g\sim 10^3$ instead of  $g\sim 1$. More than that, convexity
downwards takes place for the $\beta$-function till $g\sim 100$
\cite{4} (in the "natural" normalization) and behavior of any
quantities is indistinguishable from "trivial" in the wide range
of parameters. Nevertheless, according to \cite{4} the asymptotics
of $\beta(g)$ in  four-dimensional $\varphi^4$  theory has a form
$\beta_\infty g^\alpha$ with $\alpha\approx 1$ and the true
triviality may be absent. This point was ultimately clarified in
the paper \cite{0}.

Recent results for 2D and 3D  $\varphi^4$ theory  \cite{14,15}
also correspond to  $\alpha\approx 1$. The natural
hypothesis arises, that $\beta(g)$ has the linear asymptotics
for an arbitrary space dimension  $d$.  Analysis of
zero-dimensional theory confirms asymptotical
behavior $\beta(g)\sim g$ and reveals its origin. It
is related with unexpected circumstance that the
 strong coupling limit for the renormalized charge  $g$
 is determined not by large values of the bare
 charge  $g_0$,  but by its complex values\,\footnote{\,One can
 be anxious that the complex values of the bare charge spoils
 unitarity of theory, but this problem is easily solvable.
 One can begin with the real bare charge and prove
 unitarity of renormalized theory in the usual
 manner; it defines  theory only for $0\le g \le g_{max}$,
 where $g_{max}$ is finite. For values $g_{max}< g
 <\infty$, the theory is defined by analytic continuation, which
 conserves unitarity. In the latter case the bare charge becomes
 complex but it does not affect any observable quantities.}.
More than that, it is sufficient to consider the region $|g_0|\ll
1$, where the functional integrals can be evaluated in the
saddle-point approximation. If a proper direction in the complex
$g_0$ plane is chosen, the saddle-point contribution of the
trivial vacuum is comparable with the saddle-point contribution of
the main instanton, and a functional integral can turn to zero.
The limit $g\to\infty$ is related with a zero of a certain
functional integral and appears to be completely controllable.  As
a result, it is possible to obtain asymptotic behavior of the
$\beta$-function and anomalous dimensions: the former indeed
appears to be linear \cite{0}. Asymptotics $\beta(g)\sim g$ in
combination with non-alternating behavior of $\beta(g)$
corresponds to the second possibility in the Bogolyubov-Shirkov
classification: $g(L)$ is finite for large $L$ but unboundedly
grows at $L\to 0$. Henceforth, the true triviality of $\varphi^4$
theory is absent \cite{0}.

\vspace{3mm}
\begin{center}
{\bf 5. Numerical results.}
\end{center}
\vspace{2mm}

Existing numerical results can be divided into
several groups.
\vspace{1mm}

{\it (a) Decreasing of $g(L)$ with the growth of $L$.} Decreasing
of effective interaction  $g(L)$ was obtained in many papers (see
e.g.  \cite{32}--\cite{34}) and
indicates only that $\beta(g)$ is positive. The detailed
analysis of this decreasing can give essential
information on the  $\beta$-function, but in fact such
analysis was never performed.
\vspace{1mm}

{\it (b) RG in the real space.} This kind of research is an
approximate realization of the Kadanoff
scaling transformation  \cite{22}
in the spirit of early papers by Wilson. The system is divided
into finite blocks, which are combined thereafter into
larger blocks.  The blocks are characterized by a finite number of
parameters, whose evolution is analyzed. The papers of this
direction are characterized by high quality \cite{35,36}, but they
only demonstrate evolution of the system to the Gaussian fixed
point and confirm the initial analysis by Wilson. \vspace{1mm}

{\it (c) Logarithmic corrections to scaling.} Phase transitions
for  $d>4$ are described by the mean field theory, while for $d=4$
the corresponding power-law dependence acquire logarithmic
corrections \cite{55,16}:
$$
M \propto (-\tau)^{1/2} \left[\ln(-\tau)\right]^{3/(n+8)}\,,
$$
$$
\chi^{-1} \propto |\tau| \left[\ln|\tau|\right]^{-(n+2)/(n+8)}\,,
\eqno(7)
$$
$$
H \propto M^{3}/|\ln M|\,,\qquad \tau=0\,,
$$
etc, where  $M$, $H$, $\chi$, $\tau$ are magnetization, magnetic
field, susceptibility and the distance to the critical point in
temperature, respectively. Existence of logarithmic corrections is
beyond any doubt and their numerical verification
\cite{42}--\cite{49} is either (for $g_0\ll 1$) confirmation of
the leading logarithmic approximation \cite{55}, or (for $g_0\agt
1$) confirmation of the Wilson picture of critical phenomena.
Nevertheless, the majority of authors directly relate their
results to triviality of $\varphi^4$ theory. \vspace{1mm}

{\it (d) Extension of Eq.1 to the region of large  $g_0$.}
Dependence of the renormalized charge against the bare one for
fixed  $\Lambda/m$, studied in the papers \cite{37}--\cite{40},
looks as the only evidence of true triviality of  $\varphi^4$
theory.  The typical results of such kind \cite{37} are presented
at Fig.\,3 and indicate that dependence  $g_0$ on $L$ contains the
Landau pole ($N$ is proportional to $\Lambda/m$).
\begin{figure}
\centerline{\includegraphics[width=6.1 in]{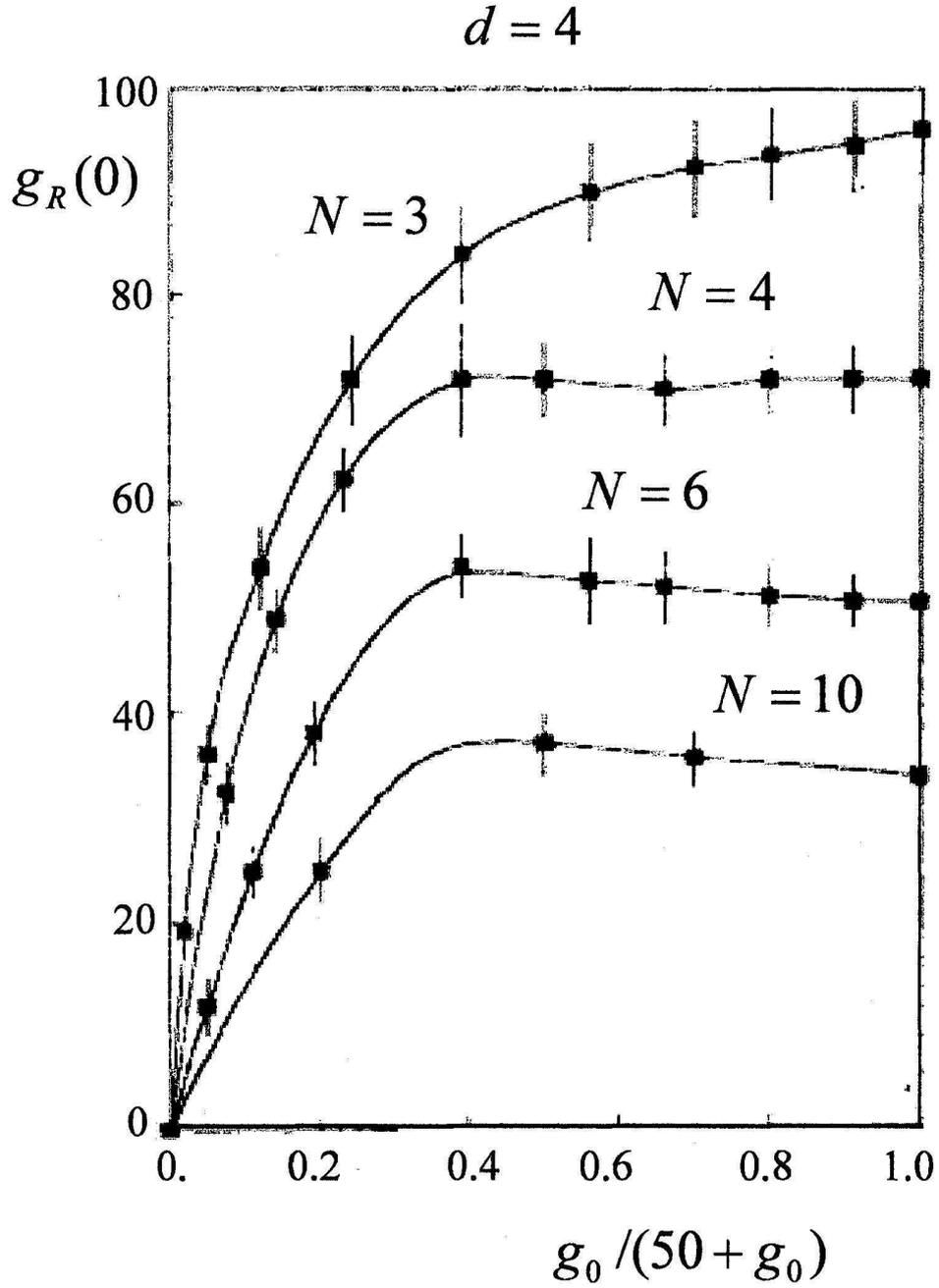}}
\caption{The renormalized charge $g_R(0)$ (estimated for zero
momenta) against the bare charge $g_0$ (corresponding to
interatomic spacing $a_0$) in four-dimensional $\varphi^4$ theory
for fixed values of $Na_0$ and $m$ but different number
$N^4$ of lattice sites
(according to \cite{37}). } \label{fig3}
\end{figure}

More close inspection reveals the typical misunderstanding related
with normalization of charge. The authors of \cite{37} were
evidently sure that values $g_0\approx 400$ lie in the deep of the
strong coupling region. In fact, all results for finite $g_0$
correspond to the parabolic portion of the $\beta$-function
(Sec.4) and do not reveal  essential deviations from Eq.1 (see a
direct comparison in \cite{38}). Only the points for $g_0=\infty$,
obtained by reduction to the Ising model, look nontrivial.
However, in the course of such reduction, the empirical dependence
$m_0^2=-const\, g_0$ (in fact, corresponding to the one-loop law)
was extrapolated to the region of large $g_0$.  Such extrapolation
is absolutely ungrounded and the results for $g_0=\infty$ are not
reliable, whereas without them no serious conclusions can be made
from Fig.\,3. Dependence $g$ on $g_0$, analogous to that in
Fig.\,3, can be obtained  also from high temperature series
\cite{40} and the lattice strong coupling expansions \cite{39};
however, these approaches also use doubtful extrapolations based
on the specific reduction to the Ising model.

In our opinion, the serious researches of such kind should
first of all reveal reliable deviations from Eq.1, related
with non-quadratic form of the  $\beta$-function. Analysis of
such deviations is the only possibility to obtain information on
behavior of  $\beta(g)$ in the strong coupling region.

The recent developments \cite{0} give new insight on the
results under discussion. Unbounded growth of $g(L)$ for
$L\to 0$ requires the use of the complex values of the bare
charge, in order to formulate the nontrivial
continuum theory.  Such possibility was not exploited in the
papers \cite{37}--\cite{40}, and their results (like Fig.\,3) do
not prove anything, even if they are taken for granted.
\vspace{2mm}

{\it (e) Papers of the recent period.}
In recent years, the aspects related with triviality are
intensively discussed in the series of papers by Agody, Consoli
et al \cite{50}--\cite{51a}. These authors suggested the
nontrivial character of the continuum limit of $\varphi^4$
theory, which constructively corresponds to rejection of the
standard perturbation expansions.

The idea is illustrated by example of non-ideal Bose gas with the
Bogolyubov spectrum  ($\epsilon(k)\sim k$ for small $k$ and
$\epsilon(k)\sim k^2$ for $k\to\infty$). The "continuum limit" of
this model can be reached by diminishing two characteristic scales
of the problem, i.e. the scattering length and the inter-particle
distance. Supporting different relationship  between two scales,
one can  either restore the quadratic spectrum of the ideal gas
("entirely trivial theory"), or obtain the strictly linear
spectrum of noninteracting phonons ("trivial theory with
nontrivial vacuum"). The latter scenario is suggested for the
continuum limit of the  $\varphi^4$ theory, in order to reconcile
spontaneous symmetry breaking with triviality.

Even if possibility of the latter scenario is accepted, the
question remains, why such scenario should be realized
physically. For example, in the case of the Bose gas of
neutral atoms, there is no real possibility to change
simultaneously both the gas density and the scattering
length. The situation suitable for the authors of
\cite{50}--\cite{51a} occurs in the case of a special
long-range  interaction, whereby a change in the density
affects the "Debye screening radius". However,
this scenario is not arbitrary and can be predicted from the
initial Hamiltonian.

According to \cite{50}--\cite{51a}, the assumption
on the nontrivial character of the continuum limit is
confirmed by  numerical modelling on the  lattice.
However, this conclusion is based not on a direct "experimental
evidence", but only on its particular interpretation.
Numerical experiments were performed deep in the region of
the one-loop law and could not contain any information
on triviality. The results, whatever unusual they
might seem, must by explained within the framework of a weak
coupling theory.

\vspace{3mm}
\begin{center}
{\bf 6. Theoretical results}
\end{center}
\vspace{2mm}

{\it (a) Arguments by Landau and Pomeranchuk.}
Landau and Pomeranchuk  \cite{3} have noticed that the
growth  of $g_0$ in Eq.1 drives the observable charge $g$
to the constant limit  $1/(\beta_2 \,\ln \Lambda/m)$,
which does not depend on $g_0$. The same behavior can be obtained
making the change of variables $\varphi\to \tilde \varphi g_0^{-1/4}$
in the functional integrals
$$
I^{(M)}_{\alpha_1\ldots \alpha_M}(x_1,\ldots, x_M)=
\int D\varphi\,
\varphi_{\alpha_1} (x_1) \varphi_{\alpha_2} (x_2) \ldots
\varphi_{\alpha_M} (x_M) \exp\left(-S\{\varphi \} \right) \,,
\eqno(8)
$$
determining the $M$--point Green functions
$G^{(M)}=I^{(M)}/I^{(0)}$, and omitting the quadratic in
$\varphi$ terms  in the action (4); then $G^{(M)}$ transfers to
$G^{(M)} g_0^{-M/4}$.  Introducing amputated vertex
$\Gamma^{(0,4)}$ by equation
$$
G^{(4)}_{\alpha \beta \gamma
\delta} = G^{(2)}_{\alpha \beta} G^{(2)}_{\gamma \delta} +
G^{(2)}_{\alpha \gamma} G^{(2)}_{\beta \delta}
+ G^{(2)}_{\alpha \delta} G^{(2)}_{\beta \gamma} -
G^{(2)}_{\alpha \alpha'} G^{(2)}_{\beta \beta'}
G^{(2)}_{\gamma \gamma'} G^{(2)}_{\delta \delta'}
\Gamma^{(0,4)}_{\alpha' \beta' \gamma' \delta'}  \,,
\eqno(9)
$$
one can see that such a change gives
$G^{(4)}/[G^{(2)}]^2=const(g_0)$, $\Gamma^{(0,4)}
[G^{(2)}]^2\propto \Gamma^{(0,4)} Z^2\propto \Gamma_R^{(0,4)}=g
=const(g_0)$, where $Z^{1/2}$ is the renormalization factor of
field $\varphi$ and notations of \cite{16,0} are used.  If
neglecting of quadratic in $\varphi$ terms
is valid already for $g_0\ll 1$, it is all the more
valid for $g_0\agt 1$:  it gives a reason to consider Eq.1 to be
valid for arbitrary $g_0$.

These considerations may appear to be qualitatively
correct\,\footnote{\,Their validity on the quantitative level is
excluded by non-quadratic form of the $\beta$-function.
In fact, the result $g =const(g_0)$ can be obtained
by the change of variables in
the functional integral only for
$g_0\gg 1$, while its validity for  $g_0\ll 1$, based
on Eq.1, may be related with other reasons; for $g_0\sim 1$
this result is probably violated but coincidence of two
constant values in the order of magnitude can be expected
from the matching condition.  } for the {\it
real } values of $g_0$, which were suggested in them.
According to \cite{0}, variation of $g_0$ along the real axis
corresponds to the change of $g$ from zero till finite value
$g_{max}$. The qualitative validity of Eq.1 for arbitrary $g_0$
 requires that  $g_{max}\to 0$ for $\Lambda\to\infty$;
the Monte Carlo results discussed above  (Fig.\,3) indicate
exactly such possibility.  To construct nontrivial theory, one
needs complex  $g_0$ with $|g_0|\alt 1$ (Sec.4): in this case one
cannot use nor discussed transformation of functional integral
(justified for $|g_0|\gg 1$), nor the formula (1). The latter is
related with the fact that perturbation  theory cannot be used
even for  $|g_0|\ll 1$, if the region is studied where instanton
contribution is essential.

\vspace{2mm}

{\it (b) Summation of perturbation series.} The first attempts
to reconstruct the Gell-Mann -- Low function by summing
the perturbation series \cite{26}--\cite{28} led to the
asymptotics  $\beta_\infty g^\alpha$ with $\alpha>1$,
showing internal inconsistency (or true triviality) of
$\varphi^4$ theory (Fig.\,2): it was one of the strongest
arguments for the corresponding time period. The different
summation result of the paper \cite{4} at least shows that
triviality cannot be reliably established from such
researches\,\footnote{\,The results of  \cite{26,27}
have the objective character and originate from
protracted  one-loop behavior of $\beta(g)$ (Sec.4).  They are
reproduced in \cite{4} as an intermediate asymptotics and can be
explained by the characteristic dip in the coefficient function.
Variational perturbation theory \cite{28} gives results close to
\cite{4} in the region $g<10$, but does not allow to obtain the
correct asymptotic behavior even theoretically.}.  On the other
hand, all results show positiveness of $\beta(g)$ and confirm
triviality in Wilson's sense.

\vspace{2mm}

{\it (c) Papers of the synthetic character.} The series of
papers \cite{52} is extensively cited as a systematic
justification of triviality of  $\varphi^4$ theory. These papers
attempt to make some kind of a synthesis of all available
information, but
contain nothing new
from viewpoint
of advancement to the strong coupling region. Conclusions made
in  \cite{52} are rather natural, since all easily
accessible information inevitably indicates
triviality due to specific features of $\beta$-function discussed
in Sec.4.

\vspace{2mm}

{\it (d) Theories with interaction $\varphi^p$.} Certain
understanding of properties of $\varphi^4$ theory can be obtained
by studing theories with more general interaction $\varphi^p$.
Consideration of the case $p=2+\delta$ with  expansion in
parameter $\delta$ gives, in the authors' opinion \cite{53}, the
serious arguments in favor of triviality.  On the other hand,
exact calculation of the  $\beta$-function in the limit
$p\to\infty$  \cite{56} gives asymptotic behavior
$\beta(g)\sim g(\ln{g})^{-\gamma}$, proving non-triviality of
theory. The latter result looks more reliable since it is not
restricted by the real values of the bare charge, which were
implicitly implied in \cite{53}.

\vspace{2mm}

{\it (e) Limit $n\to\infty$.}  The $\varphi^4$ theory is
considered to be exactly solvable in the limit $n\to\infty$
\cite{22,54}. Its $\beta$-function appears to be
effectively of the one-loop form and leads to results
like Eq.1, corresponding to asymptotics  $\beta(g)\sim g^2$.
This fact is  considered as evidence of triviality,
even in the respectful papers  \cite{54}.

In fact, coefficients of the $\beta$-function are
polynomials in $n$ and have the following structure for
$d=4-\epsilon$:
$$
\beta(g)=-\epsilon g + \beta_2 (n+a) g^2 + \beta_3(n+b) g^3
+ \beta_4 (n^2+cn+d) g^4+\ldots
\eqno(10)
$$
where $\beta_2,\,\beta_3,\,a,\, \ldots\sim 1$. The change
of variables
$$
g=\frac{\tilde g}{n}\,,\qquad \beta(g)=\frac{\tilde
\beta(\tilde g)}{n}
\eqno(11)
$$
gives
$$
\tilde\beta(\tilde g)=-\epsilon \tilde g + \beta_2 \tilde g^2 +
\frac{1}{n} f_1(\tilde g) + \frac{1}{n^2} f_2(\tilde g)
+\ldots
\eqno(12)
$$
and only two first terms remain in the  $n\to\infty$ limit. This
conclusion is valid for  $\tilde g\sim 1$ or $g\sim 1/n$,
which is sufficient for investigation of the vicinity of the
fixed point and determination of the critical exponents
\cite{22}. However, such procedure does not give any information
on the region $g\sim 1$, not to mention  $g\gg 1$. Henceforth, no
statements on triviality of $\varphi^4$ theory can be made.

\vspace{2mm}

\qquad\qquad\qquad\qquad\qquad\qquad----------------------

\vspace{2mm}

In conclusion, we have discussed the questions related with
expected triviality of four-dimensional $\varphi^4$ theory in the
continuum limit. Triviality in  Wilson's sense  is confirmed by
all available information and can be considered as firmly
established. Indications of true triviality are not numerous and
allow different interpretation.  According to the recent results,
such triviality is surely absent.

\vspace{2mm}

This work is partially supported by RFBR  (grant 06-02-17541).

\end{document}